\documentclass[journal]{IEEEtran}
\usepackage{amsmath,amssymb,amsfonts}
\usepackage{algorithm, algorithmic}
\usepackage{graphicx}
\usepackage{textcomp}
\usepackage{subfig}

\def\BibTeX{{\rm B\kern-.05em{\sc i\kern-.025em b}\kern-.08em
    T\kern-.1667em\lower.7ex\hbox{E}\kern-.125emX}}
    
\usepackage[sorting=none,backend=bibtex]{biblatex}
\begin{document}
%\\history{Date of publication xxxx 00, 0000, date of current version xxxx 00, 0000.}

\title{Shuttling in Bidimensional Segmented Ion-Trap Quantum Processors %Computers
with T-Junctions}

\author{
J.~Durandau,
C.~A.~Brunet,
F.~Schmidt-Kaler,
U.~Poschinger,
F.~Mailhot,
and Y.~Bérubé-Lauzière%
\thanks{J. Durandau, C. A. Brunet, F. Mailhot and Y. Bérubé-Lauzière are with the Institut quantique and the Département de génie électrique et de génie informatique, Université de Sherbrooke, Sherbrooke, Québec, J1K 2R1, Canada.}%
\thanks{F. Schmidt-Kaler and U. Poschinger are with QUANTUM, Institute of Physics, Johannes Gutenberg University, 55128 Mainz, Germany.}
}

%\titlepgskip=-15pt

\maketitle

\begin{abstract}

%\textcolor{red}{VERSION FSK
Shuttle-based trapped ion quantum processors typically employ a one-dimensional (1D) linear architecture to transport ion-qubits between one ore more laser interaction zones where the quantum gates are implemented, along with several qubit register storage segments. The two-dimensional (2D) quantum CCD architecture employs also T- or X-junctions for an improved scaling and efficiency. Here, we explore the shuttling layer in the compilation of quantum algorithm typical building blocks in such architecture. To weight the effort of linear shuttle and junction shuttle, we introduce individual cost functions for each operation. This allows comparing the total cost for
%genuine
quantum circuit building blocks such as the QFT, Carry, Adder, Shift, and Comparator circuits.
%, and a random sequence.
We study their scaling properties with increased qubit numbers. 
%À COÛT DE DÉPLACEMENT ÉQUIVALENT, LES ARCHITECTURES 2D TENDENT À ÊTRE SUPÉRIEURES AUX ARCHITECTURES 1D.
%JD:
At equivalent transport cost for junction and linear shuttling, we show that 2D architectures outperform the 1D linear trap with the ratio improving as the number of ions increases.
%We find, that for a cost ratio between linear shuttles and T-junction transport of 2, the 2D architecture outperforms the 1D architecture by a factor of xx to xx,
%THE RATIO SIGNIFICANTLY INCREASES FOR ARCHITECTURES WITH LARGE NUMBERS OF IONS FIG. 6
%depending on the specific buiding block. 
Finally, we discuss the use of cells, such that the entire processor is constructed from a 2D array of such interconnected cells. The work aims to optimize quantum processor architectures, implementing a co-design that fits to the specific task and scaling up in a shuttle-efficient way.
%}
\end{abstract}

%\begin{keywords}
%quantum computing, ion trap architecture, ion shuttling, initial qubit ordering, quantum gates, interaction zone
%\end{keywords}

%\titlepgskip=-15pt

\maketitle

\section{Introduction}

\IEEEPARstart{A}{\lowercase{s}}
the race for a functioning quantum computer went from searching for stable qubits to showcasing
%functioning machines,
complete operational machines,
shuttling-based trapped-ion computers with linear one-dimensional (1D) architectures~\cite{PhysRevLett.74.4091,ion_review_article,KURLEJ2024654} went ahead with stable qubits, ease of use and connectivity. Two-dimensional (2D) architectures have been  proposed~\cite{Pino2021,Scalable_ion_rf_trap}.
%They are, however, mostly theoretical and far from functional.
The approach of an X-junction in combination with a circular and linear transport region can be regarded as a transition from 1D to 2D~\cite{helios_abbrev}.
Trap junctions are complex to realize, and symmetry concerns lead to difficulties for such architectures. Thus, the most common shuttling-based trapped-ion quantum computer architecture used is still a linear segmented trap~\cite{conta2026toolchainshuttlingtrappedionqubits,hensinger_Tjct,hucul2008transportatomicionslinear,Amini_2010,Wright_2013}.

The present article uses the
%linear segmented trap ion
quantum computer architecture developed by the Quantenbit AG Schmidt-Kaler group of Mainz University as a starting point~\cite{shuttlingIonTrapQuantum}.
This architecture is a segmented Paul trap arranged in a linear fashion. Each segment of the trap can contain an ensemble of ions, hereby called a \textit{crystal}. A linear structure of crystals in such an  architecture is called a \textit{crystal chain}, which is an ordered list of crystals arranged according to their positions in the trap. Such crystals may be shuttled, or positioned at several different regions of the segmented Paul trap. One crystal is separated from the next one by a minimal distance of $\alpha$ segments owing to physical constraints.
In the framework of this paper, only one type of segment, called a laser interaction zone (LIZ), can implement the execution of quantum gates and operations for merging/splitting crystals~\cite{Kaufmann2014,Home2009,shuttlingIonTrapQuantum}.
Due to the linearity of the architecture and physical constraints that do not need to be discussed here (see~\cite{durandau2026heuristics}), crystals cannot be permuted with each other, only single ions can be exchanged between crystals through a sequence of crystal splits, crystal rotations, exchange of ions, and merge of ions to form crystals~\cite{Ruster2014}. Thus, in the course of running a quantum algorithm, sending to the LIZ two ions, which are initially separated, requires the execution of a lengthy sequence of operations, and a large number of crystals must be repeatedly moved across the LIZ~\cite{Bowler2012}. The number of these operations can be reduced when using more than one LIZ as shown in previous work~\cite{1article}. Yet, such reduction is inherently limited by the linear character of the architecture~\cite{heuristicsdurandau2026}.

In the present paper, 2D architectures are studied to overcome the limitations just discussed. The first two architectures to be discussed are linear with the addition of a T~junction. One of the arms of the T-junction will serve as a reservoir. Hence, these will be named reservoir architectures. Alternatively, a "star" architecture is investigated, in which multiple arms serve as reservoirs of ion crystals. The star architecture appears difficult to implement, but it leads to the tree architecture, an efficient and realizable architecture.

This paper is organized as follows. Section~\ref{sect:CctFitDisorg} presents metrics for studying architectures related to circuit execution, which is the purpose of a quantum computer.
Section~\ref{sect:JctCost} discusses the cost associated with a junction. Section~\ref{sect:ReservoirArchitect} presents the reservoir architecture comprising a T-junction and a single LIZ. This is followed by the description of a multi-LIZ reservoir architecture in Sect.~\ref{sect:ML}. The star and tree architectures are introduced in Sect.~\ref{sect:StarTreeArchitect}. In Sect.~\ref{sect:OutlookCellArchitect}, a future outlook is given on single-LIZ cell architectures that can be tiled on a plane. Sect.~\ref{sect:Conclu} discusses the results and compares the different architectures.
%The paper concludes with a discussion on %what a shuttled ion NISQ computer
%% with a bidimensional, or near to %bidimensional, architecture
%may look like in the near future.

\section{Circuit fit and %circuit
disorganization}
\label{sect:CctFitDisorg}

To study the implementation of a circuit and its execution on a specific architecture, two metrics will be used in this paper, namely the \textit{circuit fit} and the \textit{circuit disorganization}. 

The circuit fit, introduced in~\cite{1article}, is defined as
\begin{equation}
\mathrm{Circuit~fit} = \frac{\mathrm{Total~cost}}{\mathrm{Number~of~gates}}.
\label{eq:CircFit}
\end{equation}
%The circuit fit
It is the average cost of implementing a gate. With it, one can use the limit, as the number of ions tends to infinity, to characterize the behavior of the circuit's structure with respect to an architecture and a given shuttling algorithm. A \textit{convergent circuit fit} indicates that the circuit
%extends easily
is extensible
%scales easily
with the number of ions on a specific architecture, with the cost then becoming a multiple (proportional factor) of the circuit depth. The circuit fit is a measure of the implementation cost on a given architecture. Furthermore, in previous work~\cite{1article}, we proved the existence of a theoretical lower bound for the circuit fit. Hence, the circuit fit metric can also serve as a guide for shuttling optimization.
In this respect, the circuit fit can be specialized for a specific purpose, whereby the definition of the cost in the denominator of the circuit fit can be restricted to a part of the total cost that one wishes to target for optimization or analysis. For instance, by restricting the cost to correspond to only merge and split operations, the circuit fit will give more information on the shuttling algorithm (as these operations are more influenced by the ions' positioning in the trap), whereas displacement operations are more linked to the architecture itself (in the latter case, one would use a cost corresponding to displacement operations). 

To refine the analysis of the implementation and execution of a circuit on a specific architecture, the notion of \textit{circuit disorganization} is resorted to~\cite{heuristicsdurandau2026}.
% De manière naïve, la désorganisation est la distance entre deux ions pour lesquels il faut exécuter une porte au moment où on s'apprête à effectuer les opérations de shuttling pour exécuter la porte.
Basically, the disorganization of a circuit at a given instant is the distance between two ions on which a two-qubit gate has to be executed at the moment when the shuttling algorithm starts carrying out the operations necessary to bring the ions in the LIZ, and until the ions are brought into the LIZ.
%It measures the difference between the actual implementation of the circuit $v$ and the theoretical perfect implementation $V$ at an instant $t$. If the circuit is partitioned into its gates $g_i$, then $|V_{g_i} - v_{g_i}|$ is the disorganization of the circuit at $t$.
% METTRE REFORMULATION ICI:
More precisely, the crystal distance between two ions involved in a gate as a measure of disorganization formalizes the intuition that the more the crystals are distant, the more the architecture is disorganized with respect to the circuit to be executed. A circuit whose disorganization diminishes means that the shuttling is able to position ions closer to their ideal positioning for executing the circuit by having to move the ions as little as possible (lowest displacement cost); such a circuit is coined a \textit{convergent circuit as regards disorganization}. A circuit for which the disorganization of the ions' positioning grows as shuttling proceeds
%means that shuttling is not able to position the ions in a way that future operations will be cheap to execute; such a circuit
is \textit{divergent}.
A divergent circuit means that every added ion adds a cost to the shuttling of every other ion as the ions are positioned. %is further disorganized.
With reference to the discussion above about the circuit fit, note that a convergent circuit fit also means that the circuit disorganization is controlled. 

The two metrics just described will be used later to analyze the interplay between circuits, architectures, and shuttling algorithms.

\section{Junction cost}
\label{sect:JctCost}

Segmented traps with junctions will be considered in the sequel. A junction in such a trap is a segment where an ion can be made to change direction.
The following assumptions are made about the junction cost in this study, that is the cost related to shuttling through the junction:
\begin{itemize}
    \item The junction cost is
          %invariant for symmetric displacements.
          equal regardless to what direction the crystal is moved in the T- or X-junction.
    \item %The junction cost is invariant irrespective of  all configurations of crystal positioning.
    The junction cost is insensitive to the presence of a crystal in a junction segment not involved in a displacement.
    \item The junction cost is comparable to the displacement cost between linear segments.
\end{itemize}

\begin{figure*}[htb!]
\centering
\includegraphics[width=0.8\textwidth]{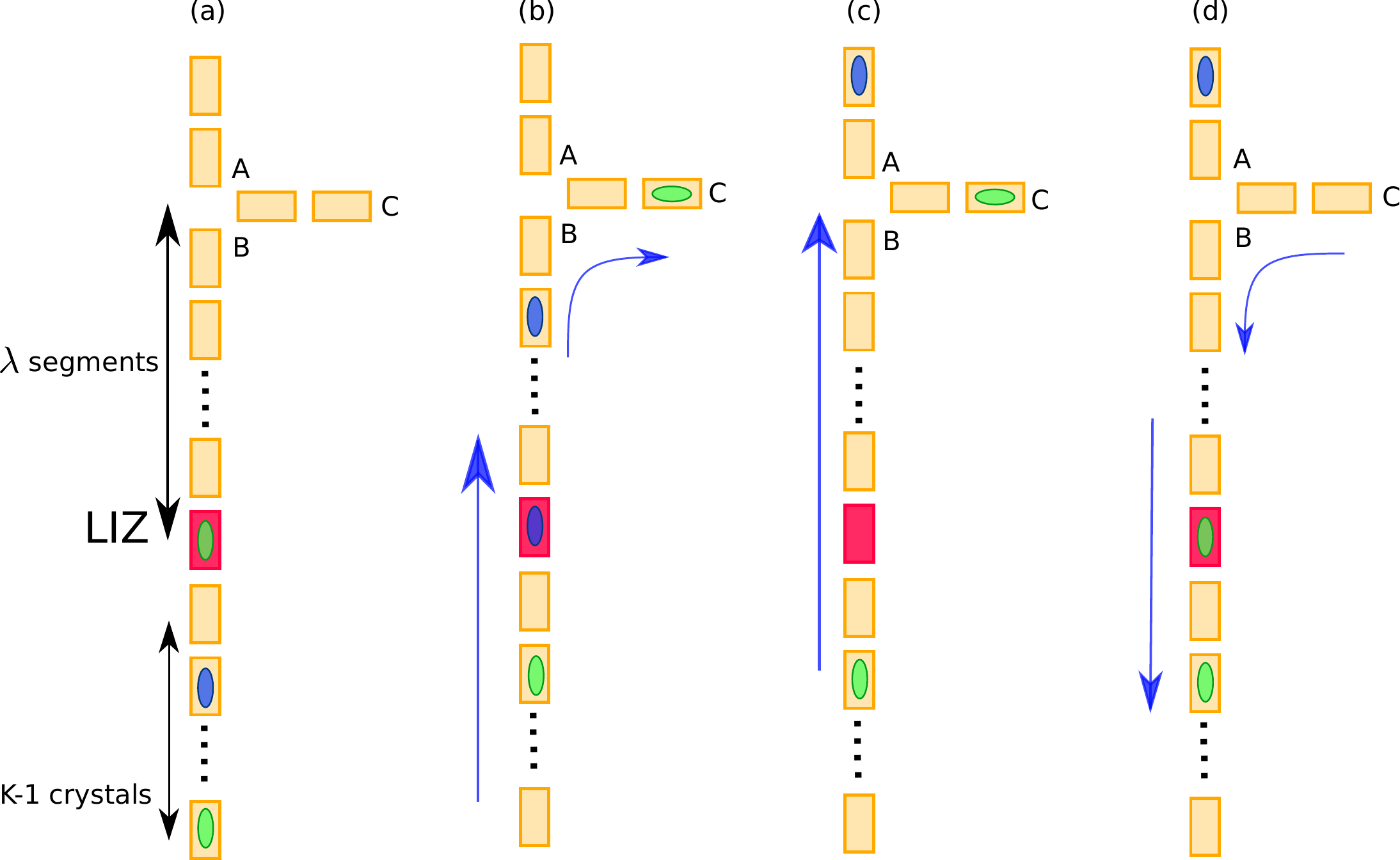}
\caption{Illustration of the exchange of ions in a reservoir architecture between the two green crystals with $K-1$ crystals overall present in the trap. (a) Initial position of crystals. (b) The top green crystal is sent to branch~C of the reservoir; the rest of the crystal chain moves up to just below the reservoir in branch B. (c) The $K-2$ blue crystals are sent to branch A and the bottom green crystal is sent to the LIZ. (d) The top green crystal is sent to branch B and then to the LIZ for ion swap. Both crystals are now adjacent, the junction cost is $(K-2)C_{AB} + 2C_{BC}$.}
\label{fig:ReservoirArchi}
\end{figure*}

The first assumption means that the displacements A-B and B-A between segments A and B, as shown in Fig.~\ref{fig:ReservoirArchi}, are equivalent in cost and time~\cite{Amini_2010}. It also means that the displacements A-C and B-C are equivalent. Furthermore A-C and C-A are equivalent. In other words, the junction cost is invariant in all directions. This assumption follows from the symmetry of the junction and holds in typical experimental situations.

% Reduire le nombre de point, reduire les segments, reduire la figure

The second assumption means that if segment A is occupied by a crystal, the cost of displacement through B-C is not affected by the presence of a crystal in segment A.
In practice, however, it may be necessary, on physical grounds, to leave segment A empty when a B-C displacement is executed.
% FORMULATION ORIGINALE DE FSK: This reduces the crosstalk of control voltages required for the T-junction transport of one crystal on the other crystal which is stored in the reservoir.
This reduces the crosstalk on a crystal, which is stored in the reservoir, of control voltages required for the transport of another crystal through the T-junction.
Such an empty segment when operating a junction displacement may cause the time implementation to grow as the crystals must be shuttled serially to have such an empty segment. This does not affect, however, the behavior of the shuttling algorithm as this does not alter the topology, nor the shuttling strategy.

Finally, the last assumption means that the cost (\textit{e.g.} $C_{AB}$) of a displacement across a junction is of the same order as other costs encountered in a linear 1D trap. More precisely, the cost of shuttling through a junction will be considered to have a value between that of the cost of a linear displacement across a segment and that of a split or merge operation~\cite{Heating_rate} (highest cost - worst case). It cannot yet be predicted what values junction costs will have, but a bound will be given where the use of junctions becomes advantageous. In what follows, the previous assumptions will be considered to hold.
We argue that in general, the shuttle time in a linear segment with respect to a T- or X-junction will be proportional as the mass and charge of the ion and the range of the voltages will determine this in combination with cleverly designed electrode geometry.   

The common ion order (CIO) shuttling algorithm for a unidimensional architecture has been described in past work~\cite{heuristicsdurandau2026}. As the best shuttling algorithm, it is used as a baseline to compare with. The present article will refer to the unidimensional architecture using the CIO algorithm as CIO-Uni.

\begin{figure}[ht!]
\centering
\includegraphics[width=0.4\textwidth]{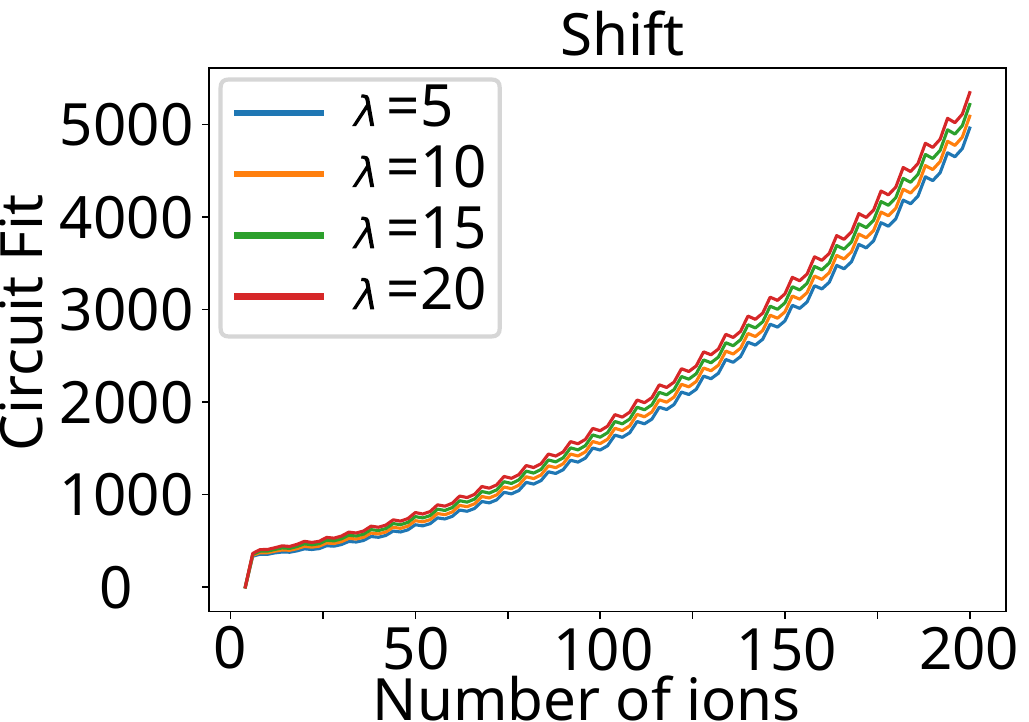}
\caption{Circuit fit for a Shift circuit as a function of the number of ions for different values of $\lambda$ (distance between LIZ and junction).}
\label{fig:lambdaDiff}
\end{figure}

\section{Reservoir architecture}
\label{sect:ReservoirArchitect}

The reservoir architecture has a uni-LIZ linear configuration with one segment replaced by a T-junction at a distance of $\lambda$ segments from the LIZ, see Fig.~\ref{fig:ReservoirArchi}. The main interest here is that ions can be shuttled through the crystal chain without having to resort to costly ion exchanges. This results in a significant reduction in the number of merge and split operations as the need to exchange ions occurs only when the ions are in different crystals and only once per gate.
Hence, in this architecture the contribution of merges and splits in the circuit fit is bounded between 3 and 6. This shows the superiority of the bidimensional case with T-junctions, since in the previous uni-dimensional case, the cost of merges and splits was a major contribution of the circuit fit, whereas here it is only very minor.

The performance of this reservoir architecture will be compared with that of a linear trap in conjunction with the CIO algorithm presented previously~\cite{1article,heuristicsdurandau2026}.

The displacement cost of the worst case for implementing a two-qubit gate in a linear uni-LIZ architecture is~\cite{1article}
\begin{equation}
\alpha K(K-1) + P(K-1),
\label{eq:UniCost}
\end{equation}
where $K$ is the number of crystals, $\alpha$ the minimal number of segments between crystals imposed by the trap physics ($\alpha = 1$ for the Mainz "red" trap architecture~\cite{Hilder2022}), and $P$ is the cost of one ion exchange. This value will serve as a reference.
Using the entire shuttling of a worst case in the reservoir architecture presented in Fig.~\ref{fig:ReservoirArchi}, the displacement cost $K(K-1)$ becomes $\alpha K(\frac{\lambda}{\alpha} + K-1)$ since the distance between the LIZ and the junction $\lambda$ is added. As there is only one ion exchange, a cost of only one $P$ remains ($P(K-1)$ becomes $P$). Finally, the cost of using the junction is $(K-2)C_{AB} + 2C_{BC}$.
Thus, for the worst case where two ions are at opposite ends of the crystal chain, the total cost is given by
\begin{equation}
\alpha K(\frac{\lambda}{\alpha} + K-1) + (K-2)C_{AB} + 2C_{BC} + P.
\label{eq:worstCostReservoir}
\end{equation}
This cost only makes sense in the case where crystal permutations must take place. Thus, at least three crystals ($K>2$) are necessary for this formula to hold; any smaller number of crystals will not use any of the advantages given by the architecture.

% LE DÉBUT DE CE PARAGRAPHE (DEUX PREMIÈRES PHRASES) FONT SUITE À UN COMMENTAIRE DE FERDINAND QUI DEMANDAIT D'AJOUTER CE TEXTE.
% LA PREMIÈRE PHRASE DE FERDINAND ÉTAIT PLUTÔT FORMULÉE COMME SUIT (JE L'AI SIMPLIFIÉE):We have compiled the circuits of basic and frequently used building blocks for quantum circuits, which are the quantum Fourier transform (QFT), sequences for carry and shift, and the circuits for a comparator and an adder.
We have compiled the circuits of basic and frequently used building blocks for quantum circuits, which are the quantum Fourier transform (QFT), Carry, Shift, Comparator and Adder. We are interested in such building blocks, as they rely on different qubit connectivities, which in turn results in a variation of the costs. 
Fig.~\ref{fig:lambdaDiff} shows, with an illustrative example using the Shift circuit, the dependence of the circuit fit on the distance $\lambda$ between the LIZ and the junction. It is seen that the circuit fit is steadily higher the larger $\lambda$ is. This shows that the LIZ and the junction should be as close as possible. Since the junction may have an undesirable influence on the LIZ, a non-vanishing distance $\lambda$ may be necessary for the physical implementation. Since the displacement cost across one segment is considered to be 1 (this is a convenient normalization we chose), $\lambda$ is also the cost of displacement across $\lambda$ segments.

Note that in general, the cost of a displacement over a distance of $\lambda$ segments may be less than $\lambda$, because the acceleration at the start and the end are the challenging parts of a displacement. In this paper, we still assume a stepwise displacement  between segments as this may be easier to calibrate and scale to large devices.

Writing $C_{BC}$ as a multiple of $C_{AB}$ as $ C_{BC} = \xi C_{AB}$ with $\xi \in \mathbb{R}^+$; the cost function given in Eq.~\eqref{eq:worstCostReservoir} then becomes
\begin{equation}
    \alpha K(\frac{\lambda}{\alpha} + K-1) + C_{AB}(K-2+2\xi) + P.
\label{eq:worstCostReservoirVariant}
\end{equation}
In order that the architecture with a reservoir be advantageous over that of a linear uni-LIZ, one obtains from Eqs.~\eqref{eq:UniCost} and~\eqref{eq:worstCostReservoirVariant} the following inequality:
\begin{equation}
\begin{split}
    \alpha K( \frac{\lambda}{\alpha} + K -1) + C_{AB}(K-2+2n) + P
    \\
    < \alpha K(K-1)+ P(K-1).
\end{split}
\end{equation}
This leads to the following condition on the cost for which the worst case for a reservoir architecture is better:
\begin{equation}
    C_{AB} < \frac{P(K-2)-\lambda K}{K-2+2\xi}.
\end{equation}
For large $K$, the bound on $C_{AB}$ becomes
\begin{equation}
C_{AB} < P-\lambda .
\label{eq:limitCost}
\end{equation}
It can be deduced from this equation that $\lambda$ shall not be too large for this architecture to be of interest. This is a strong constraint of the reservoir architecture, which is of interest for implementation as this trap topology is worth building only if this relation is satisfied. 

\begin{figure*}[ht!]
\centering
    \subfloat[\label{subfig:QFTResGraph}]{
      \includegraphics[width=0.3\linewidth]{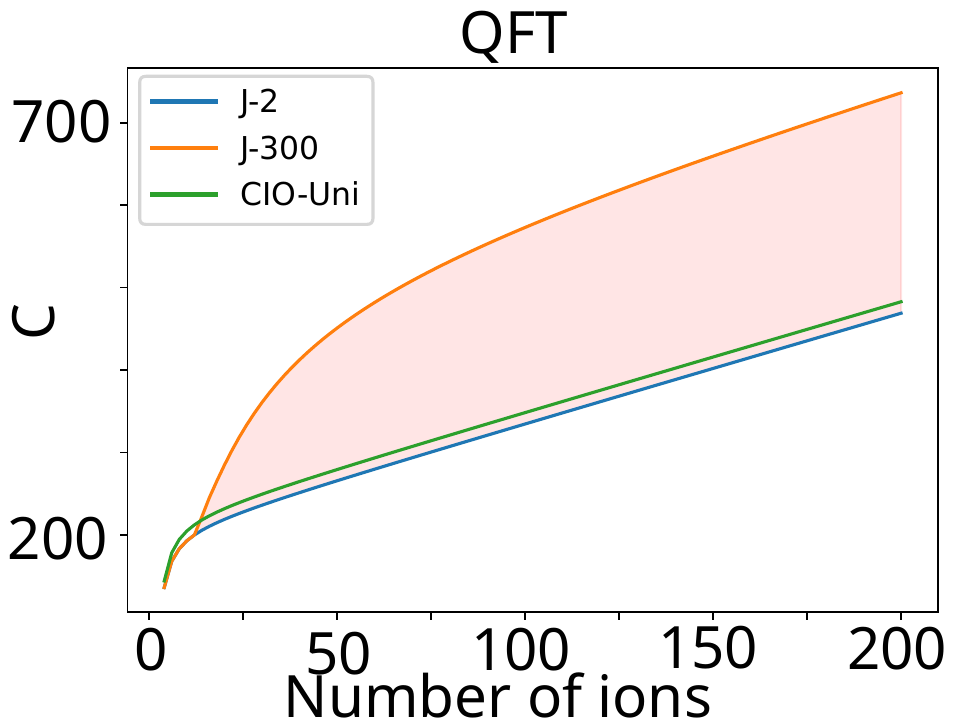}
    } 
    \subfloat[\label{subfig:CarryResGraph}]{
      \includegraphics[width=0.3\linewidth]{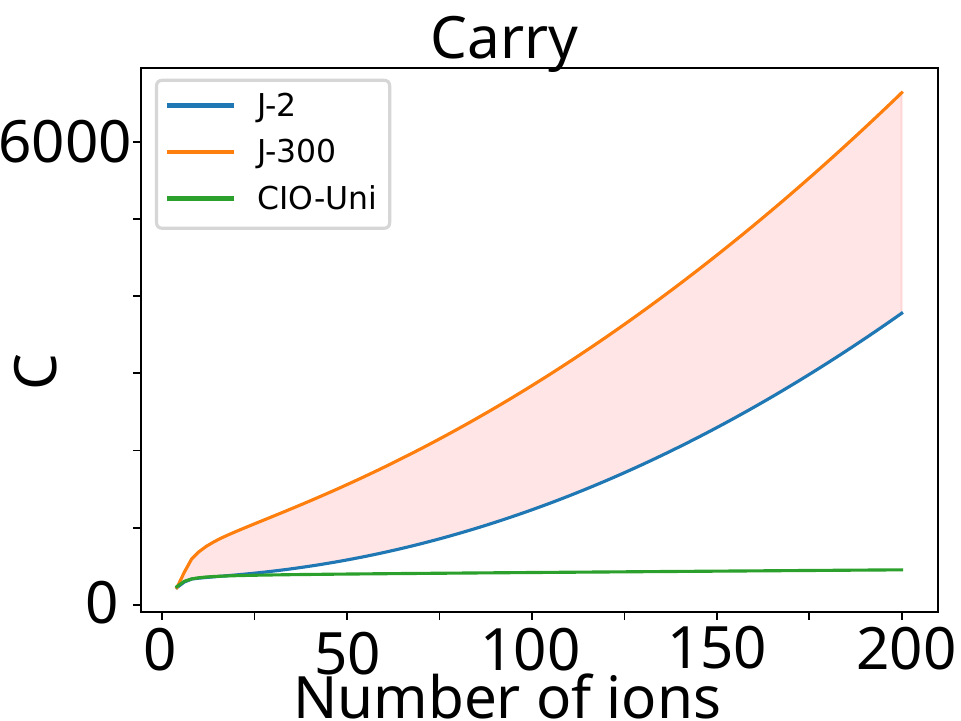}
    } 
    \subfloat[\label{subfig:ComparatorResGraph}]{
      \includegraphics[width=0.3\linewidth]{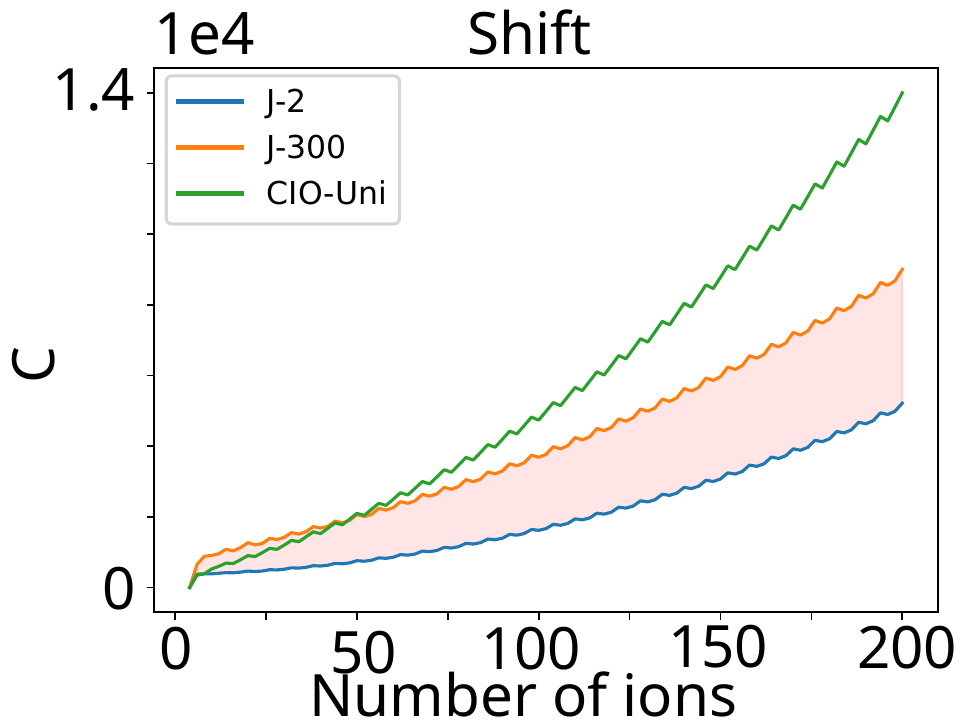}
    } \\
    \subfloat[\label{subfig:ShiftResGraph}]{
      \includegraphics[width=0.3\linewidth]{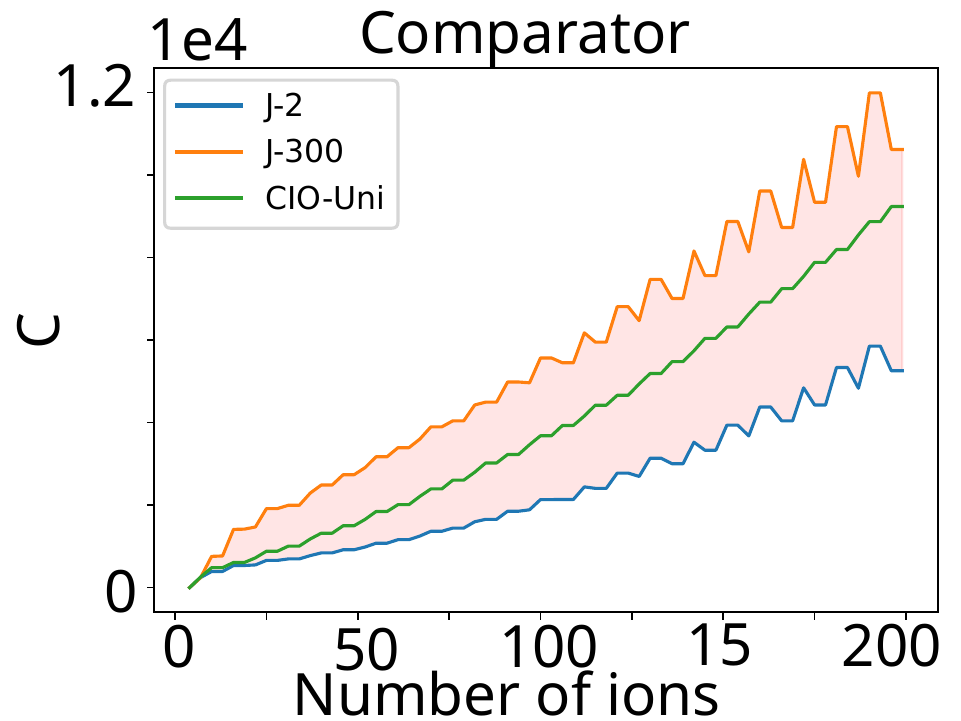}
    }
    \subfloat[\label{subfig:AdderResGraph}]{
      \includegraphics[width=0.3\linewidth]{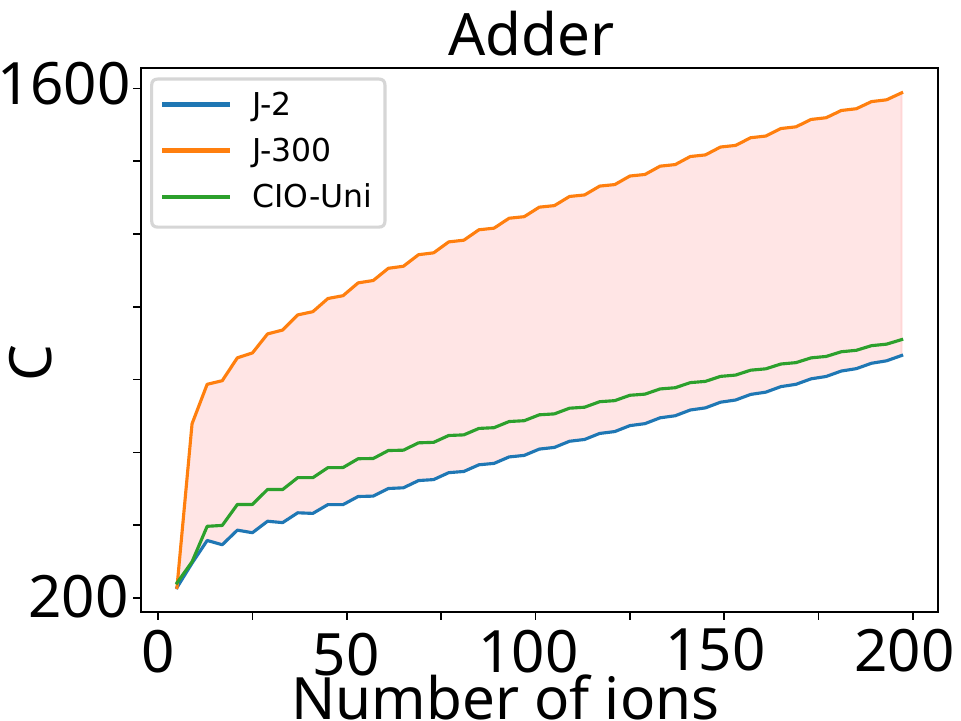}
    }

    \caption{Simulation results for the circuit fit in a reservoir architecture with $\lambda =15$ for (a)~the QFT, (b)~Carry, (c)~Shift, (d)~Adder, and (e)~Comparator circuits. The green curve CIO-Uni is the circuit fit for a linear uni-LIZ architecture, serving as reference. J-2 and J-300 are bounding curves respectively corresponding to a junction cost of 2 and 300.}
    \label{fig:reservoirResultat}
\end{figure*}

% Aggrandir bidi & uni dans la meme colonne de texte
% Change unidimensional par CIO-Uni
% CHange bidimensional par reservoir
\begin{figure}[h]
\centering
    \subfloat{
      \includegraphics[width=0.4\textwidth]{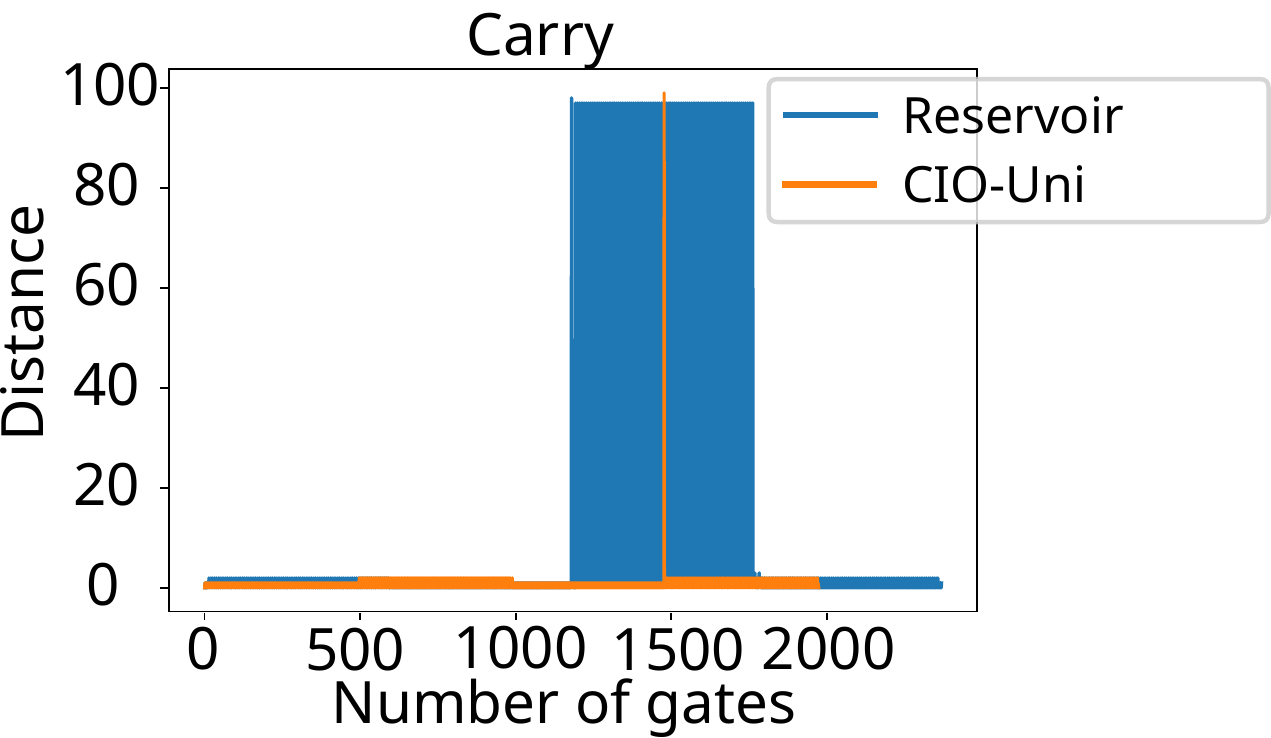}
    } 
    %\\
    %\subfloat{
   %   \includegraphics[width=0.4\textwidth]{AdderDistanceInnerGraph.pdf}
    %} 
    \caption{Comparison of the evolution of the circuit disorganization as shuttling proceeds in CIO-Uni and Reservoir architectures for
    %the Carry and~(b) Adder circuits.
    the Carry circuit.
    } 
    \label{fig:ReservorInner}
\end{figure}

Fig.~\ref{fig:reservoirResultat} shows simulation results comparing the CIO-Uni with the reservoir architecture for a junction cost ranging from 2 (denoted J-2) to 300 (denoted J-300) - here T-junctions are considered, but Y-junctions could also be considered; the reasoning would be the same.

Superior performance is obtained with the reservoir architecture for almost all circuits in the case of small junction costs (exemplified by J-2), except for Carry for which the cost of the CIO-Uni is lower than for the reservoir architecture for all cost figures (J-2 to J-300). A small improvement is seen for the QFT with J-2 (the J-2 curve being below that of the CIO-Uni). In the case of the QFT no large improvement is expected since the CIO-Uni is very well-adapted to the QFT. For the Shift circuit, the improvement is significant especially at high numbers of ions, similarly for the Comparator circuit, but to a lesser extent. For the Adder circuit (Fig.~\ref{fig:reservoirResultat}~(e)), J-2 is superior to CIO-Uni, but the two seem to converge to the same value for large numbers of ions.
To understand the result for Carry, which may at first appear surprising in comparison to other circuits, one must investigate the shuttling itself in terms of disorganization (distance). This is shown in Fig.~\ref{fig:ReservorInner}, which depicts the evolution of the distance between crystals that must be made adjacent to implement gates.
%The result for the Carry circuit (Fig.~\ref{fig:ReservorInner}~(a)) shows that this is the only circuit for which the reservoir architecture is worse. 
It is seen that this is caused by a greater disorganization in the shuttling, because, while a linear architecture only permits ion exchanges, the algorithm for the reservoir architecture is forced, if it can, to only use crystal exchanges (and not ion exchanges). In a linear architecture, only the ion of interest is moved during shuttling, while in the reservoir architecture, the targeted ion is moved as part of a crystal, hence alongside a passenger ion, and it is this passenger that causes disorganization since it should not be moved. Of course, any algorithm having Carry as one of its building blocks would be executed in a linear part of the architecture.
It may be concluded here, that the reservoir architecture
%, and thus more generally bi-dimensional topologies,
can
%generate shuttling with
execute algorithms at
smaller costs, yet the circuit disorganization becomes greater as the number of shuttling degrees of freedom becomes larger, which can be detrimental, as the Carry circuit shows.

As the results in Fig.~\ref{fig:reservoirResultat} show, for some circuits such as the
QFT %(Fig.~\ref{fig:reservoirResultat}~(a)) % YBL: ON DIT DÉJÀ AU DÉBUT DE CE PARAGRAPHE QU'ON RÉFÈRE À LA FIGURE; JE NE PENSE PAS NÉCESSAIRE DE DIRE À QUELS GRAPHIQUES DANS CETTE FIGURE ON RÉFÈRE; LE LECTEUR N'A QU'À REGARDER.
and
Adder %(Fig.~\ref{fig:reservoirResultat}~(e))
circuits, the reservoir architecture is better only for a small junction cost (close to J-2). For other more complex circuits, such as the Shift %(Fig.~\ref{fig:reservoirResultat}~(c))
and Comparator %(Fig.~\ref{fig:reservoirResultat}~(d))
circuits, the reservoir architecture offers a much larger range of in which the cost is reduced in comparison to the CIO-Uni.
It may thus be concluded that a reservoir architecture is preferable for complex circuits.

\section{Multi-LIZ architecture}
\label{sect:ML}

\begin{figure}[t!]
\centering
\includegraphics[width=0.48\textwidth]{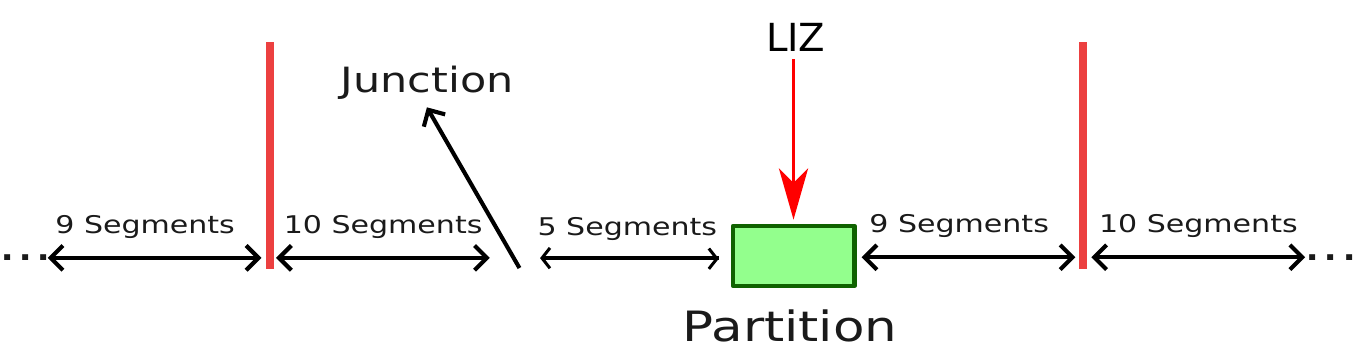}
\caption{Schematic of part of a multi-LIZ reservoir architecture. See text for description.}
\label{fig:multi_archi}
\end{figure}

To further
inquire about %study
the reservoir concept, a multi-LIZ reservoir (mLIZ-reservoir) architecture is investigated, see Fig.~\ref{fig:multi_archi}.
%It consists of a set of partitions of the whole trap
The whole trap is subdivided into a set of partitions, with
each partition having 24 basic segments with a LIZ at segment 10 and a junction at segment 16 for a total of 26 segments.
In Fig.~\ref{fig:multi_archi}, only one partition delimited by the red vertical bars is fully illustrated. Within a partition, the LIZ and the junction are separated by $5$ segments; the LIZ is 9 segments away from its closest end (to the right) of the partition and the junction is 10 segments away from its closest end (to the left), but otherwise these values have been arbitrarily chosen.
In this mLIZ-reservoir architecture, two consecutive LIZ are separated from each other by a distance of $25$ segments (including the junction).
Each partition is repeated 50 times. Using the same machine constraint as before (one segment free between each crystal), there can be up to $k = 12$ crystals per partition ($k$ and $K$ shall not be confused, $K$ was the total number of ions in the reservoir architecture with one LIZ, whereas here $k$ is the number of ions per partition, hence per LIZ).

\begin{figure*}[thp]
\centering
    \subfloat[\label{subfig:QFTResMLGraph}]{
      \includegraphics[width=0.3\linewidth]{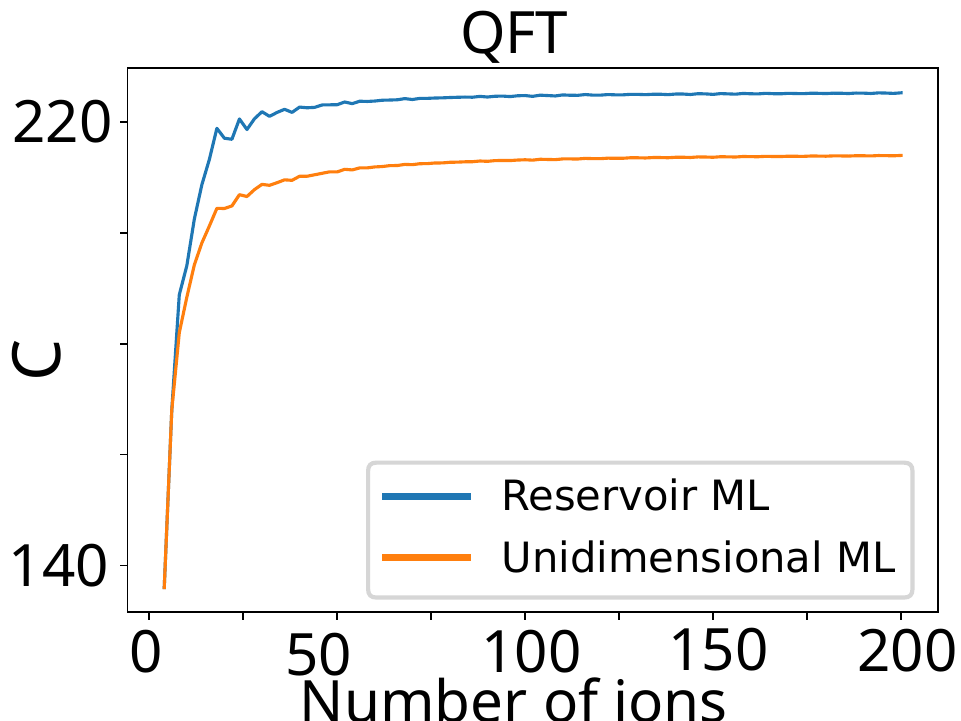}
    } 
    \subfloat[\label{subfig:CarryResMLGraph}]{
      \includegraphics[width=0.3\linewidth]{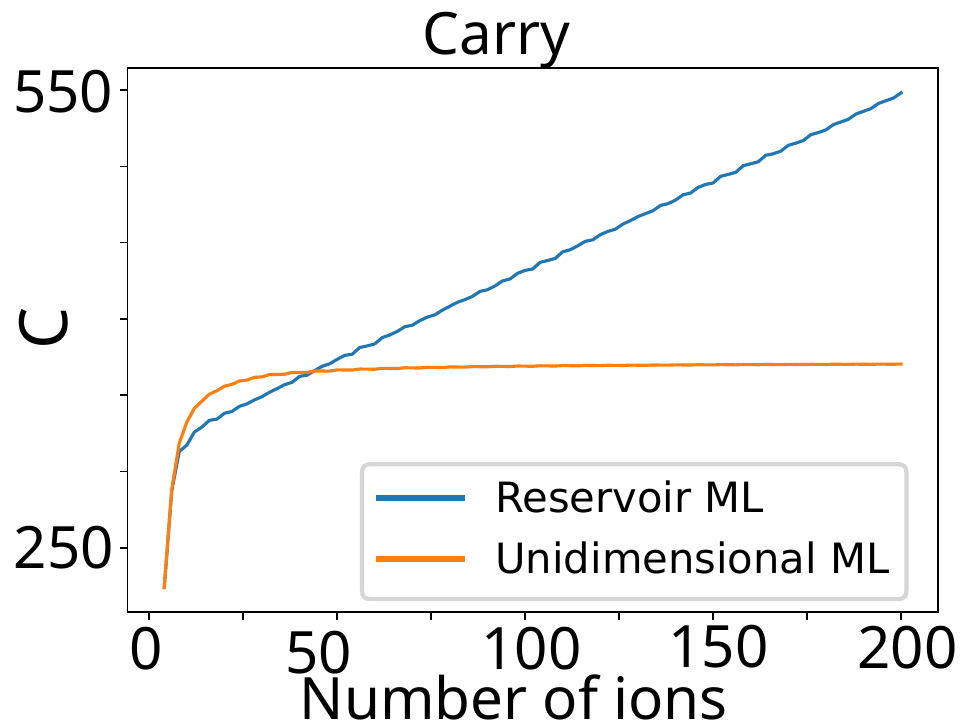}
    } \subfloat[\label{subfig:ShiftResMLGraph}]{
      \includegraphics[width=0.3\linewidth]{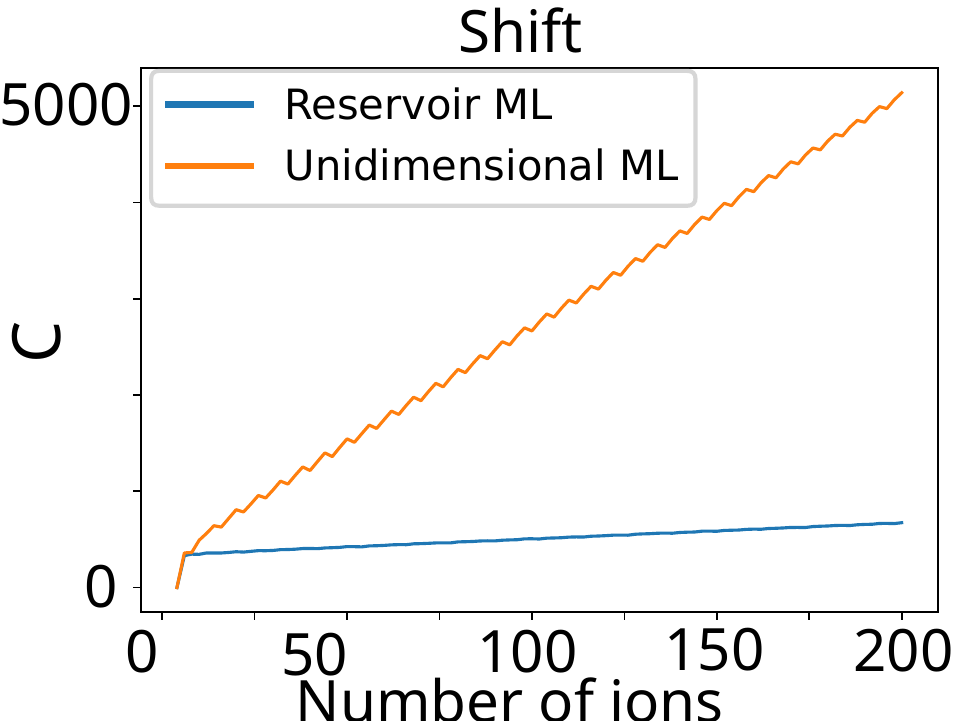}
    }\\
    \subfloat[\label{subfig:ComparatorResMLGraph}]{
      \includegraphics[width=0.3\linewidth]{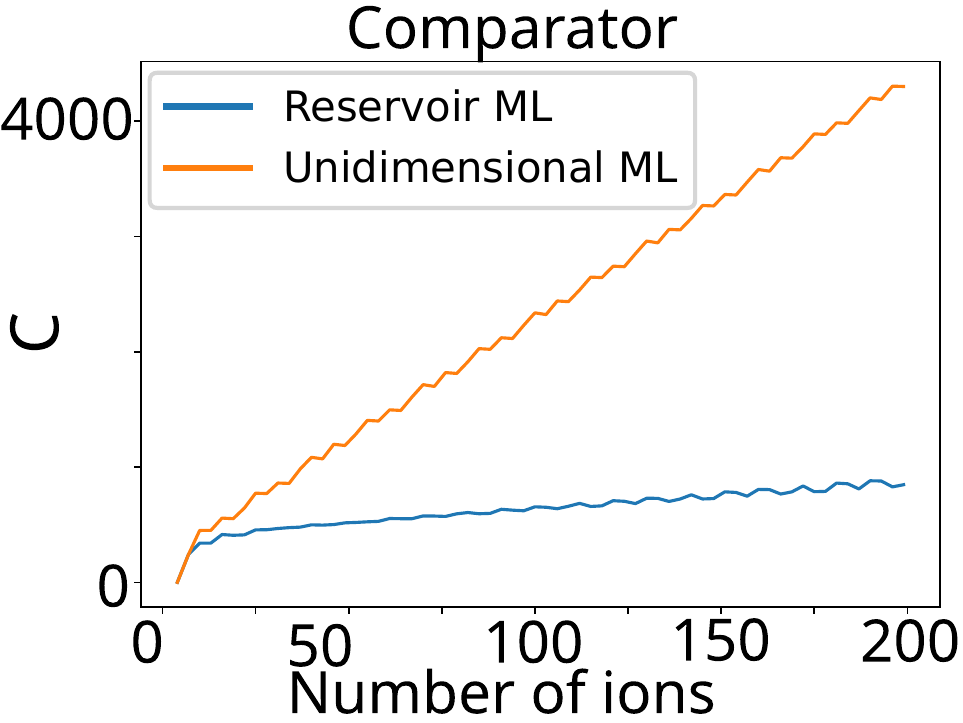}
    }
    \subfloat[\label{subfig:AdderResMLGraph}]{
      \includegraphics[width=0.3\linewidth]{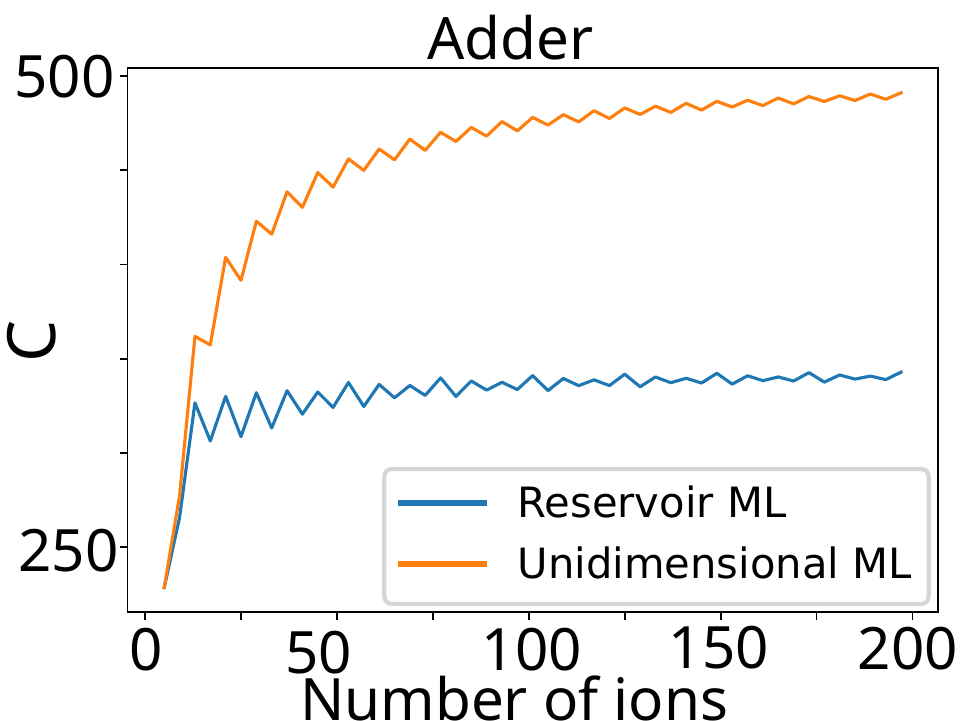}
    }

    \caption{Circuit fit for a multi-LIZ reservoir architecture with $\lambda = 5$, and $k = 12$ compared with a linear multi-LIZ architecture (without reservoir) with $k = 3$. Note that for a linear multi-LIZ architecture, the smaller $k$ is, the more efficient is the architecture, hence the choice for $k=3$.}    
    \label{fig:reservoirMLResultat}
\end{figure*}

Results for the mLIZ-reservoir architecture are presented in Fig.~\ref{fig:reservoirMLResultat} in comparison with a linear unidimensional multi-LIZ architecture with no reservoir (unidim-mLIZ). Since the distance between two LIZ is smaller for the unidim-mLiz, the QFT is less costly. The disorganization for the Carry circuit still persists. The mLIZ-reservoir is shown to perform better for all other circuits. Thus, the reduction of displacements is preserved with an mLIZ-reservoir architecture.

The addition of reservoirs to a linear architecture gives the capability to greatly improve the efficiency of ion shuttling. It allows the efficient implementation of a greater number of circuits. This is limited by the cost of junctions and the additional disorganization as the shuttling proceeds because of passenger ions as discussed previously.  
Better algorithms are needed to fully use the capability of the reservoir architecture.
Furthermore, the LIZ-junction distance $\lambda$ may become a limiting factor depending on the minimal LIZ-junction distance.
This can be illustrated with an example. Considering an ion exchange cost $P=300$ and
say that a T-junction with 
a junction cost $C_{AB} = 250 $ can be engineered, it is found according to Eq.~\eqref{eq:limitCost}, that $\lambda$ is limited to 50 segments in order that the architecture be of interest.

%Reduire taille, réduire n branches tailles
\section{Star and tree architectures}
\label{sect:StarTreeArchitect}

\begin{figure}[htb]
\centering
\includegraphics[width=0.5\textwidth]{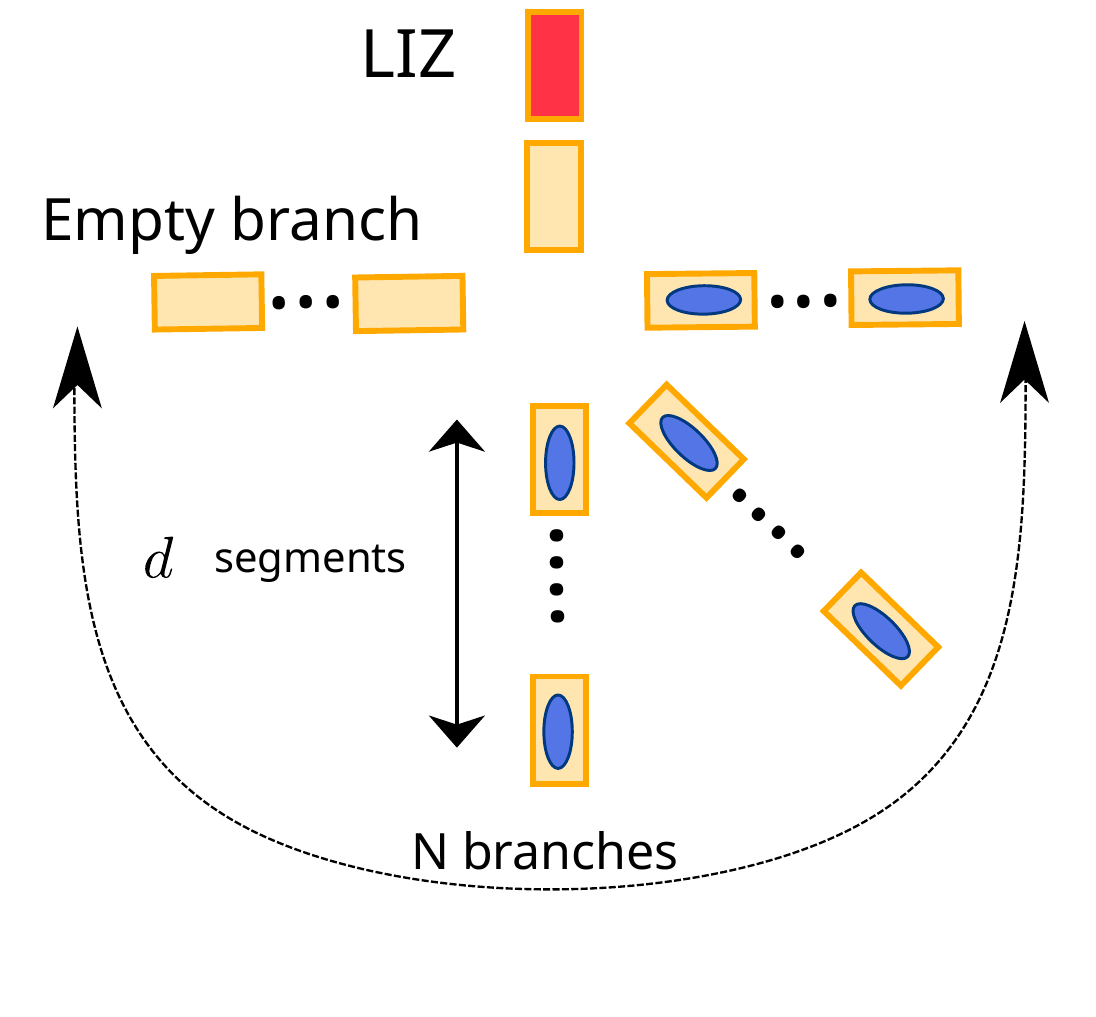}
\caption{Star architecture with $N$ branches of depth $d$.}
\label{fig:StarArchi}
\end{figure}

The star architecture is shown in Fig.~\ref{fig:StarArchi}.
% FERDINAND A DEMANDÉ À ENLEVER CELA (IL DIT: you dont know, we possibly could do this!); J'AI REFORMULÉ POUR LAISSER LA PORTE OUVERT QUE ÇA SERAIT POSSIBLE.
%It is theoretical in that it cannot be physically realized for a large number of ions. Nevertheless, the
This architecture may seem to be only of theoretical interest, as it appears % somewhat
%difficult in practice
challenging as of now
to build the central junction for a large $N$.
%number of branches (\textit{e.g.} for a large number of ions).
%a its heart
%connecting several branches.
%as the simulation use up to two hundred branches (one branch every 1.8°), one cannot physically have a junction with so many branches.
Nevertheless, the
objective of presenting this architecture is threefold: First to introduce a way to reduce the number of ion exchanges to one per gate at worst by way of a bi-dimensional architecture, second to show an architecture that bounds the maximal cost of implementing a gate, and third it leads to the tree architecture which is more easily amenable to implementation.

As seen in Fig.~\ref{fig:StarArchi}, the star architecture
%comprises $N$ branches, each with $d$ segments. Each segment can contain up to $K_d$ crystals.
is composed of $N$ branches, each
%with
comprising
$d$ segments, with $d$ being a fixed architecture constant. Each of these branches can contain up to $K_d$ crystals. There is also one LIZ branch that possesses just enough segments to implement ion exchanges and other crystal operations. All branches join at one end at a central junction to form a star. Increasing the number of ions requires adding more branches to the junction.

\begin{figure}[t]
\centering
\includegraphics[width=0.3\textwidth]{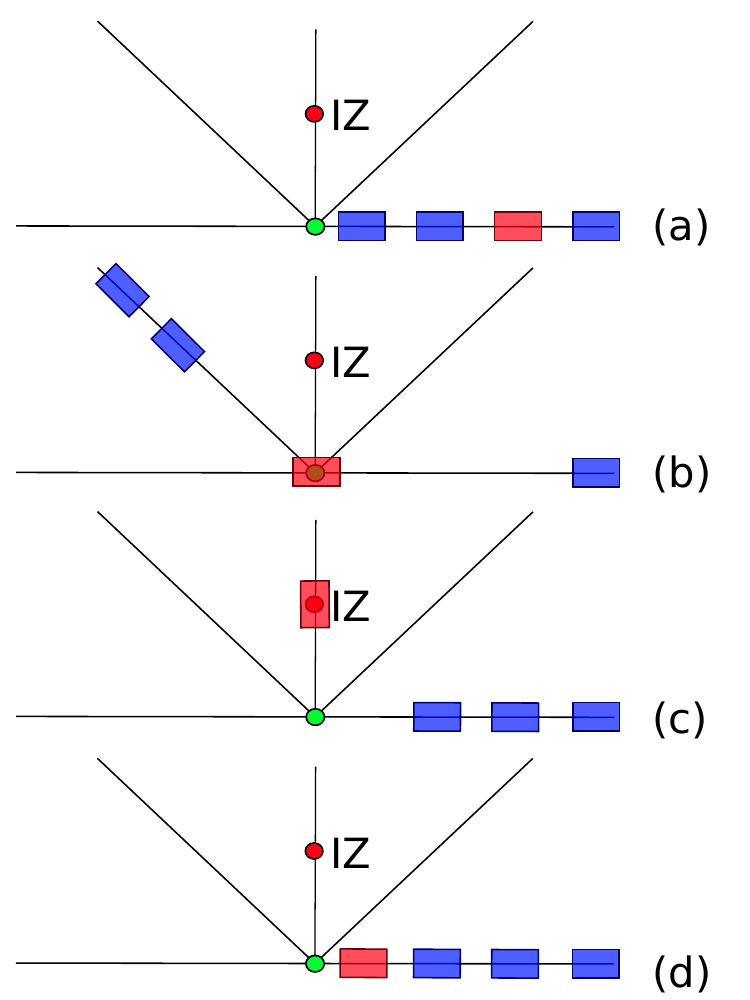}
\caption{Schematic of the shuttling algorithm to send the red crystal in the LIZ to implement a gate: (a) state before shuttling; (b) the branch of the red crystal is emptied until the red crystal is at the junction; (c) the red crystal is sent to the LIZ to implement the gate while the blue crystals are sent back in their original branch; (d) the red crystal is sent back to its branche. In case of a gate needing two different crystals, the (a) to (c) parts of the algorithm are carried out on the second crystal after the (c) part of the first crystal, and then (d) part is performed on the two crystals.}
\label{fig:StarArchiAlgo}
\end{figure}
%Uniformisé a b c & d entre ()

\begin{figure*}[htbp]
\centering
    \subfloat[\label{subfig:QFTStarGraph}]{
      \includegraphics[width=0.3\linewidth]{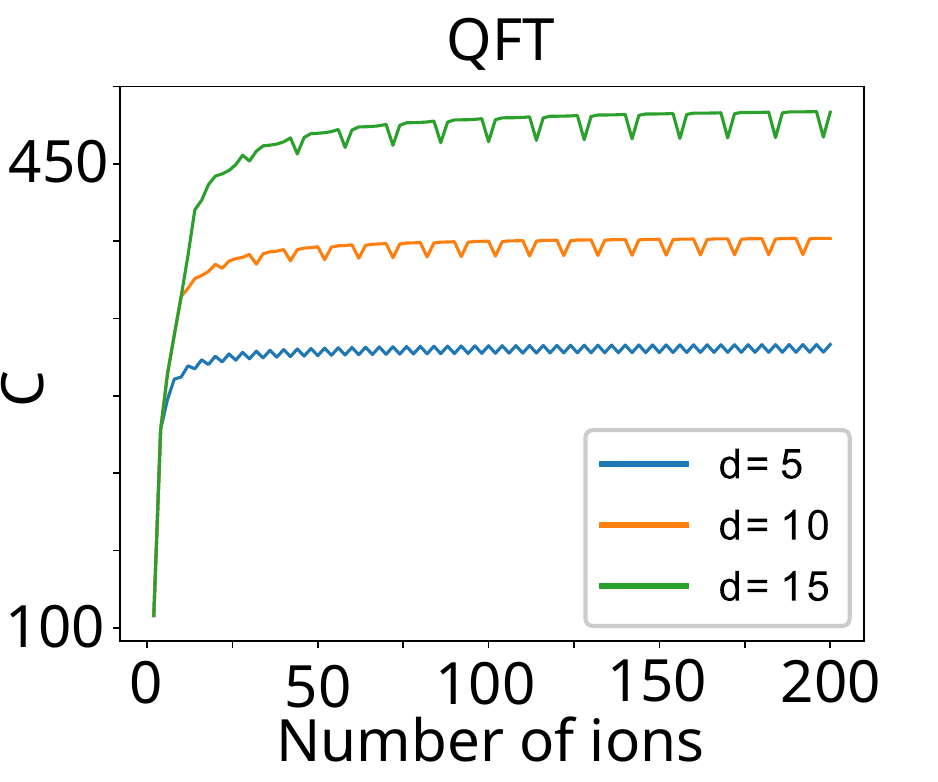}} 
    \subfloat[\label{subfig:CarryStarGraph}]{
      \includegraphics[width=0.3\linewidth]{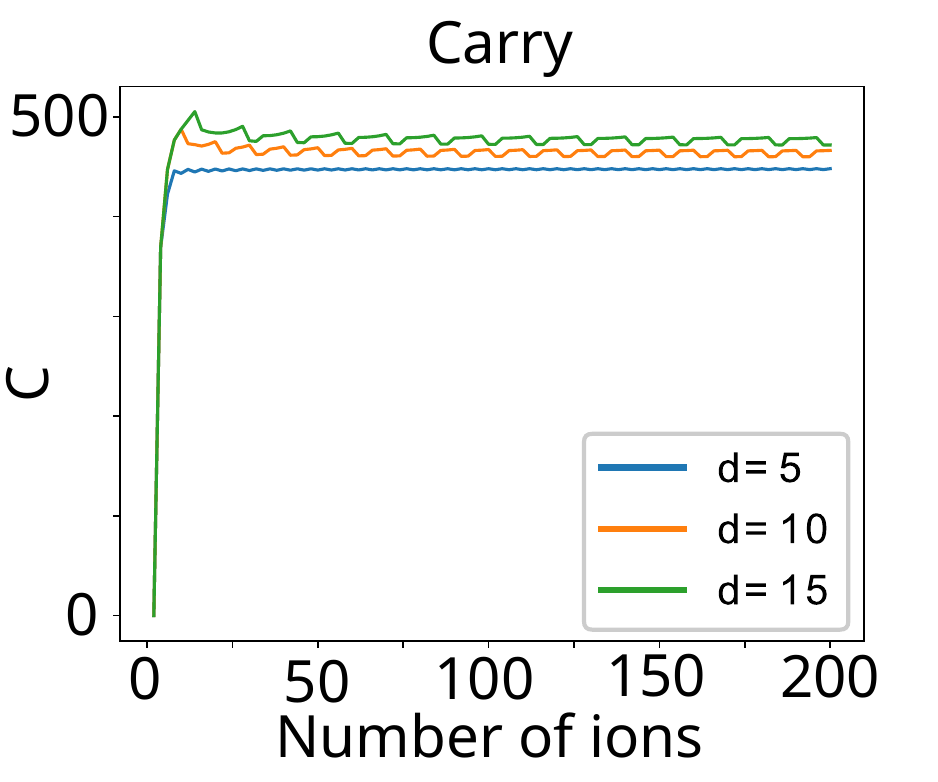}} 
    \subfloat[\label{subfig:ShiftStarGraph}]{
      \includegraphics[width=0.3\linewidth]{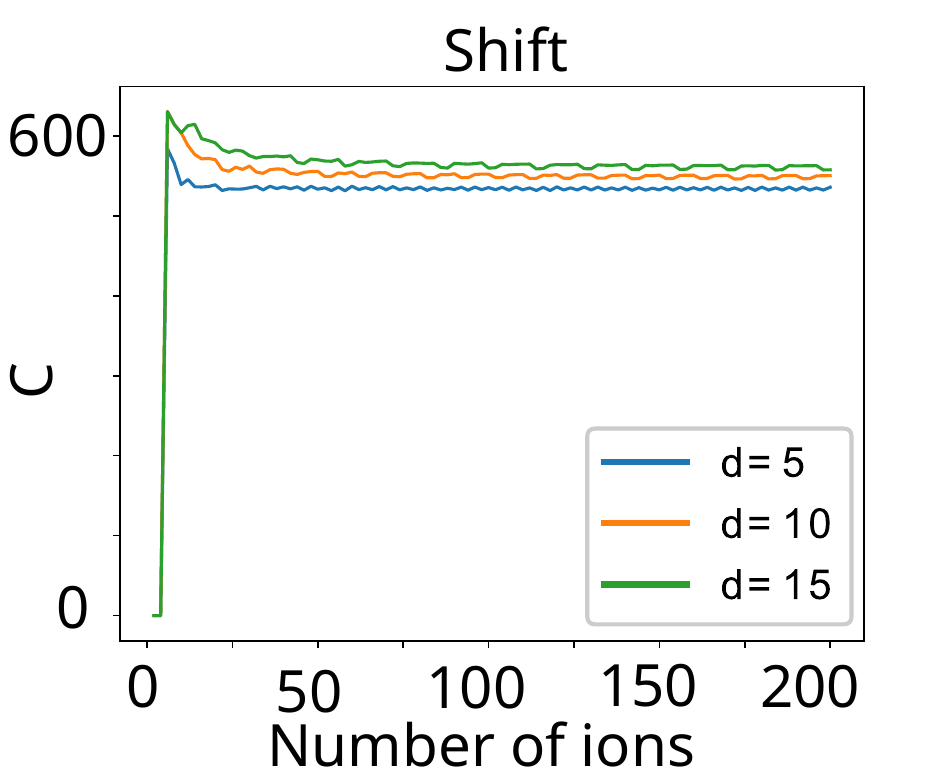}}
    \\
    \subfloat[\label{subfig:AdderStarGraph}]{
      \includegraphics[width=0.3\linewidth]{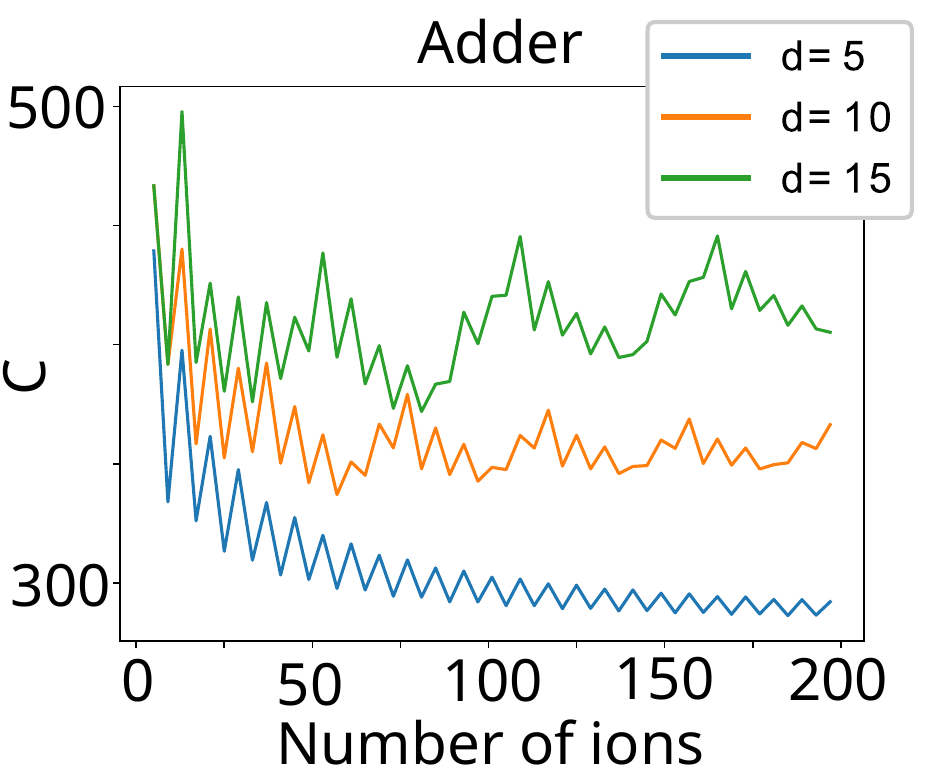}}
    \subfloat[\label{subfig:ComparatorStarGraph}]{
      \includegraphics[width=0.3\linewidth]{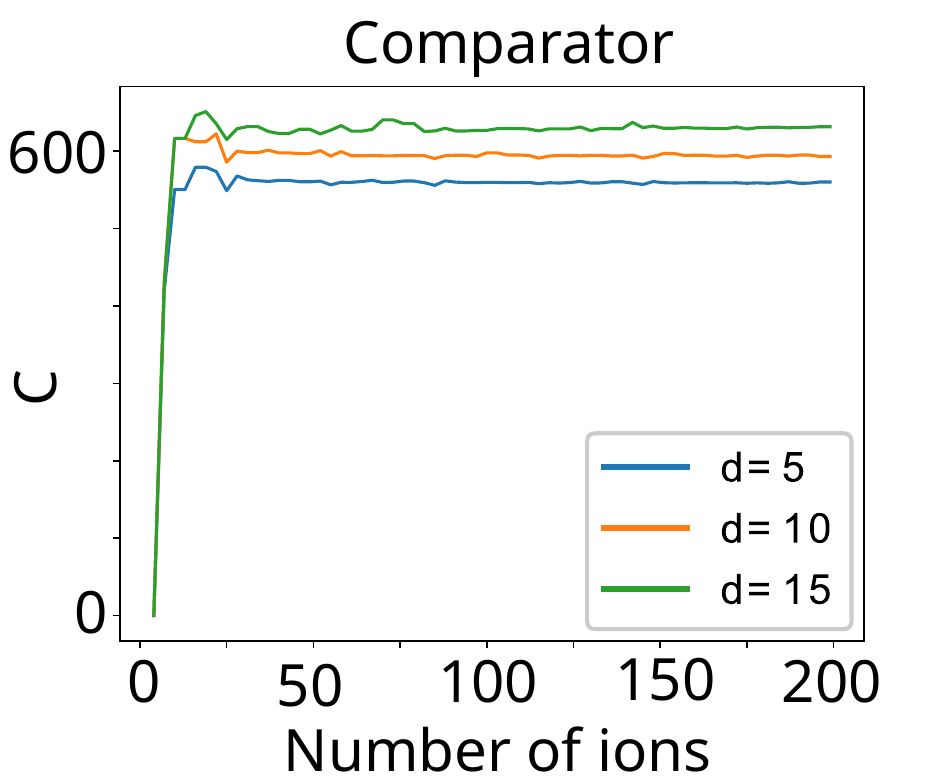}}
    \caption{Simulation results for the circuit fit for star architectures with depth $d = 5$, 10, and 15.}
    \label{fig:ResultStar}
\end{figure*}
%Size Y go to 600 each

\begin{figure}[t]
\centering
\includegraphics[width=0.25\textwidth]{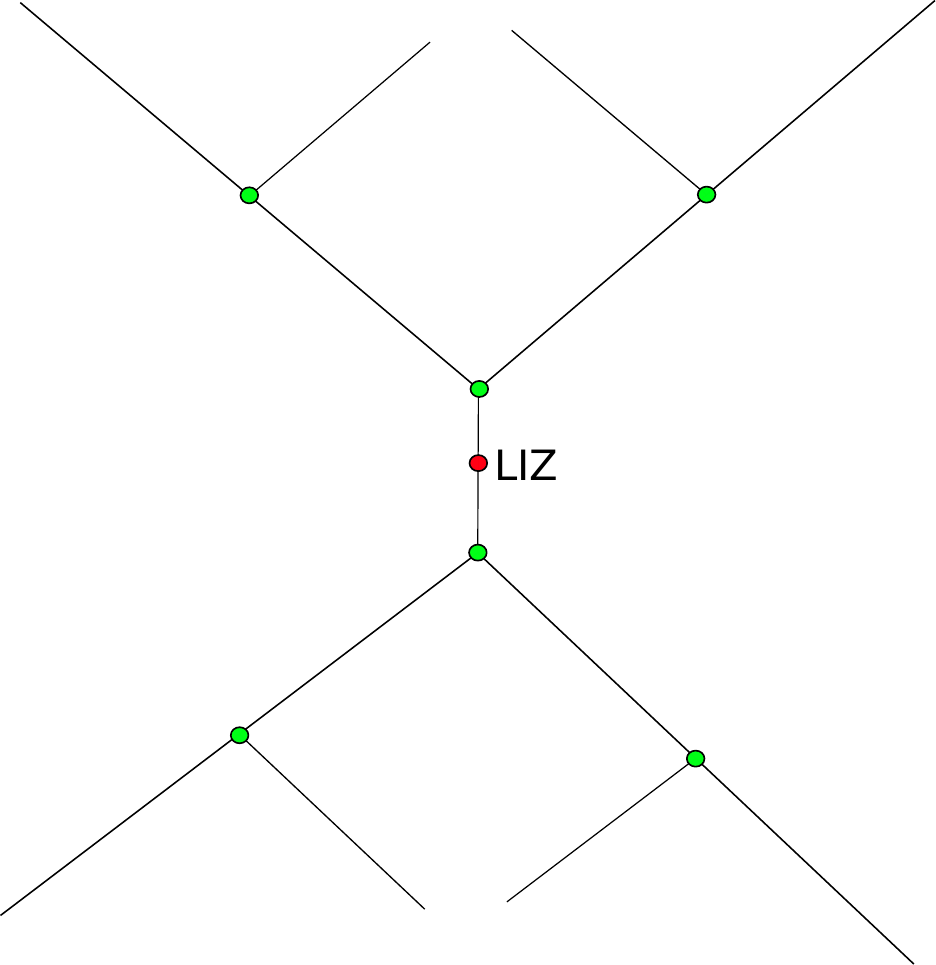}
\caption{Tree architecture using Y-junctions.}
\label{fig:TreeArchi}
\end{figure}

The architecture can contain $(N-1) K_d$ ions and not $N K_d$ as could be expected, since one branch is purposely kept empty (the empty branch) for shuttling reasons as will be seen shortly. The shuttling algorithm used here for this architecture is simple and proceeds as follows (Fig.~\ref{fig:StarArchiAlgo}). To implement a gate, the crystals containing the targeted ions are first sent to the LIZ. To move a crystal to the LIZ, the branch containing the targeted crystal is moved segment-wise to the empty branch up until the targeted crystal is in the junction, at which point it is diverted to the LIZ branch. Then the crystals in the empty branch are sent back to their original branch. If two crystals are needed for the gate, then the procedure is repeated for the second crystal. Once the gate is executed, the crystals are sent back to their branches. The underlying idea of this architecture is readily understood: From the perspective of the branch containing a crystal of interest for a gate, that branch and the empty branch are essentially equivalent to a linear uni-LIZ architecture. % UNE QUESTION QUI ME VIENT: POURQUOI N'A-T-ON PAS ÉTUDIÉ UNE ARCHITECTURE UNI-LIZ À RÉSERVOIR DANS LAQUELLE LA LIZ EST DANS LE RÉSERVOIR ? ON POURRAIT ESSENTIELLEMENT FAIRE CE QU'ON FAIT ICI DANS L'ARCHITECTURE EN ÉTOILE.

The shuttling cost is now analyzed. First, because the bi-dimensionality removes the need for most of the ion exchanges performed for shuttling as compared to a uni-LIZ linear architecture~\cite{1article},
at most one ion exchange is needed, which contributes a cost of 6 merge/split operations (see again~\cite{1article}).
This means that displacements become the only operations contributing to the cost that may grow uncontrollably in the architecture.
Yet, for the star architecture presented, the displacement cost %circuit fit
is also bound, as will now be demonstrated.

First consider the worst case: Two crystals must go to the LIZ, and each is positioned at the end of different branches.
To empty one branch entirely, each crystal must travel $d + \alpha$ segments, with each crystal traversing the junction once. Considering the cost of a displacement through the junction to be $J_c$, the cost to empty the branch is:
\begin{equation}
(K_d-1)(d+\alpha + J_c).
\end{equation} 
The displacement cost of a target crystal to the LIZ with the LIZ at a distance $Z_c$ to the junction is thus: 
\begin{equation}
   d + J_c + Z_c .
\end{equation}

The cost of displacement to the LIZ must be counted twice, because two crystals are brought to the LIZ. Once the gate is executed, these crystals must be brought back to their branches, giving the following cost per crystal:
\begin{equation}
\alpha+ J_c + Z_c.
\label{worst_case_star}
\end{equation} 
The total cost of the gate is thus
\begin{align}
  C_{\mathrm{tot},\mathrm{gate}}
    &= 2(K_d-1)(d + \alpha + J_c) \\
    &+ 2(d + J_c + Z_c) + 2(\alpha + J_c + Z_c) . \notag
\end{align}
Since $K_d = \frac{d}{\alpha}$, the last result becomes
%
%\[ 2(\frac{d^2}{\alpha} + d(1+\frac{J_c}{\alpha} -d-\alpha - J_c) + 2(d+ J_c + Z_c) + 2(\alpha+ J_c + Z_c), \]
\begin{equation}
  C_{\mathrm{tot},\mathrm{gate},\mathrm{max}} =
  2\frac{d^2}{\alpha} + 2d(1+\frac{J_c}{\alpha}) + 4Z_c + 2J_c .
\end{equation}
In this expression for the worst-case cost of a gate, only the constants $d$, $\alpha$, $Z_c$, and $J_c$ appear, which are fixed by the architecture. This means that the implementation cost of one gate is bounded. The upper bound of the circuit implementation cost is given by the product of the number of gates in the circuit and the worst-case cost. % SUPERBE DÉMONSTRATION  JONATHAN ! ;-)

Fig.~\ref{fig:ResultStar} shows simulation results for the star architecture. In all cases, the circuit fit is bounded, which supports the previous mathematical demonstration; note that as $d$ decreases, the cost decreases. Furthermore, it is seen in Fig.~\ref{fig:ResultStar}~(d) that the Adder is somewhat unstable, yet a reduction of almost half the cost is observed in passing from $d = 15$ to $d = 5$; this is also observed for the QFT (Fig.~\ref{fig:ResultStar}~(a)).
Notice that the circuit fit displays
%an oscillatory behavior
oscillations. These originate in the filling of a branch. This is caused by the choice of considering the worst case, whereby whole branches are filled as much as possible.

These results demonstrate the existence of a theoretical architecture that bounds any circuit implementation.
As mentioned, the star architecture is of theoretical interest, but it is possible to arrive at a topology that would in practice approximate it.
This is the tree architecture shown in Fig.~\ref{fig:TreeArchi}, which retains some of the advantages of the star architecture. In the tree architecture, the addition of ions does not require the addition of branches to the central junction, but rather of Y-junctions at the ends of already established branches. This causes the fixed $d$ of the star architecture to be replaced by $\ln(g(K))$, with $g$ being a function that indicates the depth of the tree branches as a function of the number of crystals. As the branch depth $d$ is no longer constant, this architecture
%still
gives a logarithmic cost growth in the worst case. This is the most advanced uni-LIZ architecture we believe possible in the short term.

\begin{figure}[t]
\centering
\includegraphics[width=0.5\textwidth]{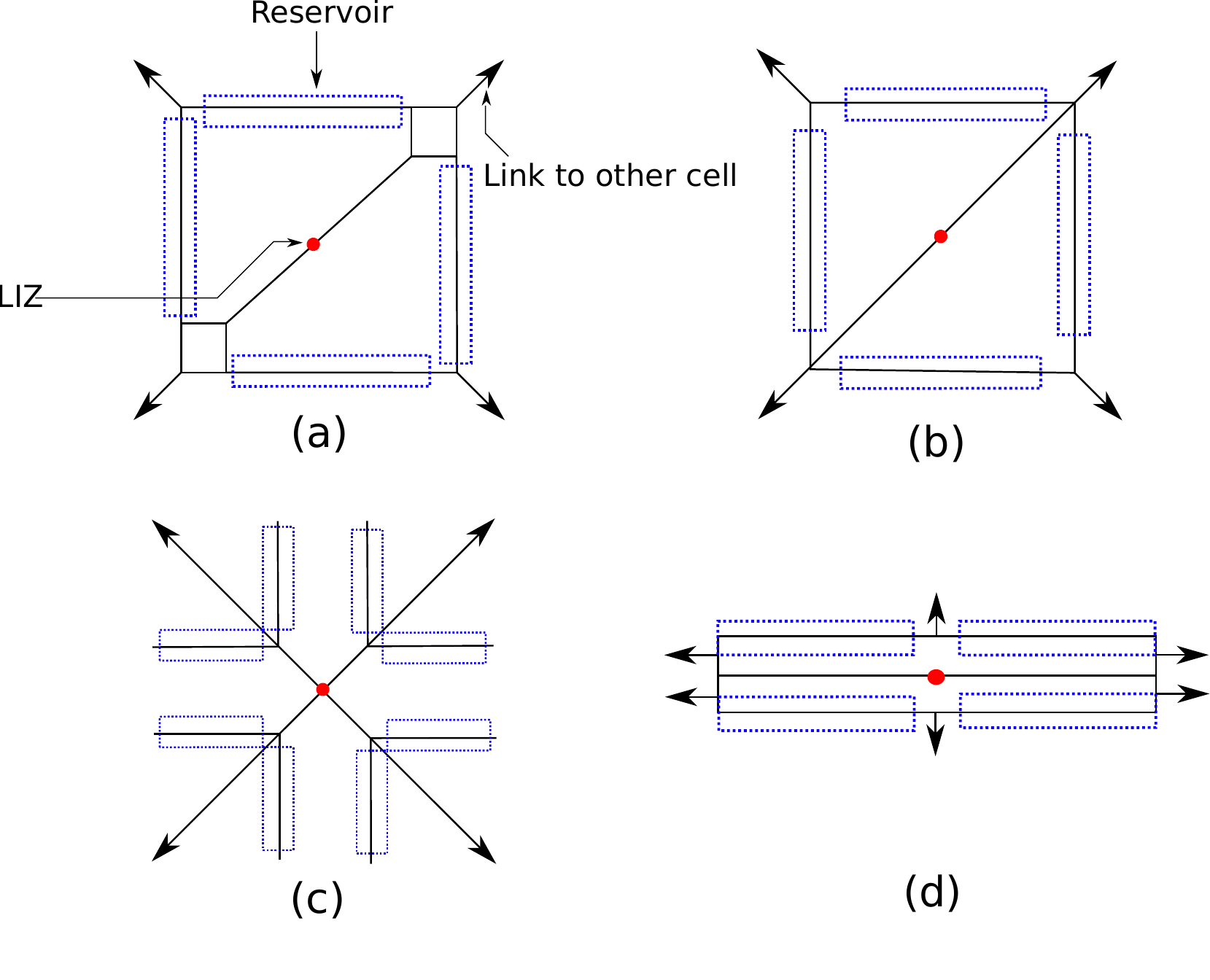}
\caption{Cell architectures. A red dot indicates the position of a LIZ. The blue rectangles show the position of reservoirs that can hold crystals not involved in a gate to be executed. The arrows show connection points with adjacent cells. The cell in: (a) uses Y-junctions, (b) X-junctions, (c) X-junction with a special LIZ-junction, and (d) T-junction.}
\label{fig:CellArchi}
\end{figure}

\begin{figure}[h]
\centering
\includegraphics[width=0.5\textwidth]{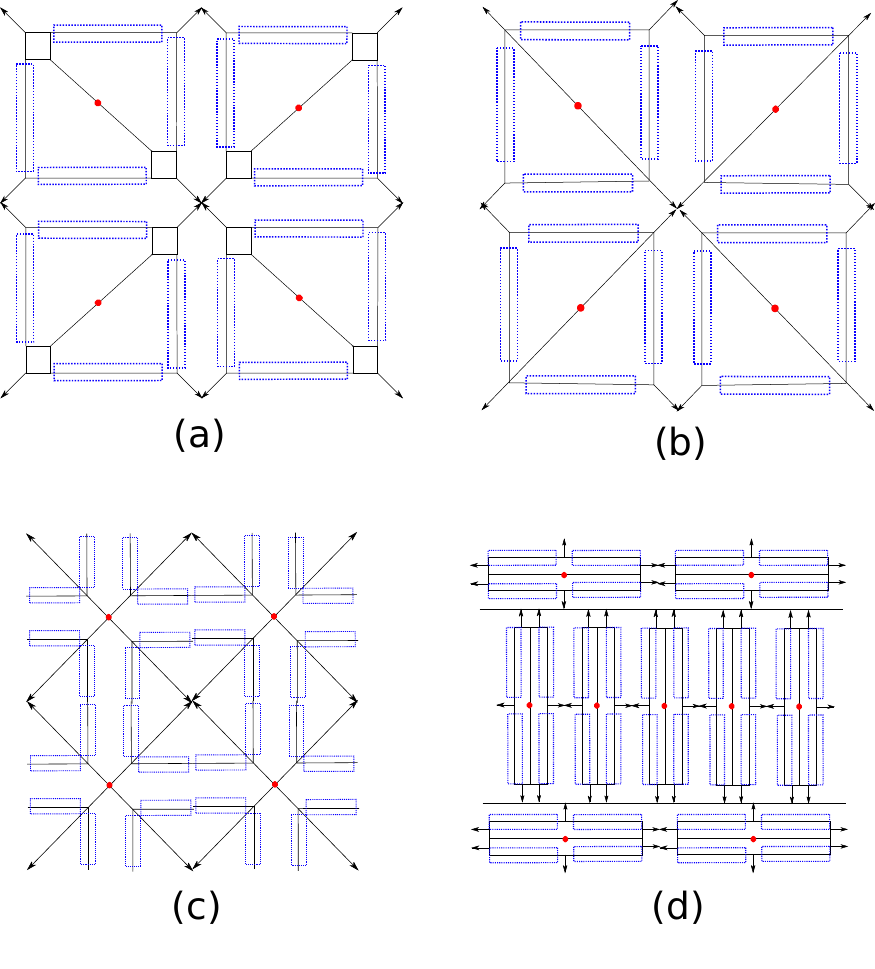}
\caption{Tiling of the plane with cells.}
\label{fig:CellArchiPatern}
\end{figure}

\section{Future outlook - Cell architectures} % A proposal
\label{sect:OutlookCellArchitect}

This section presents uni-LIZ cell architectures with reservoirs that can be easily repeated on a plane. Such 2D cells are presented as a possible area of future research. 
The main objective is to create a minimal cell that can be tiled over a planar area for a generic scaling to large  processors. This results in a small architecture that can be extended to a computer capable of implementing a large number of gates in parallel. Here are the guiding design principles: 
\begin{itemize}
    \item The cell must have the smallest possible area to maximize the number of LIZ.
    \item The cell must be symmetric around the LIZ.
    \item There must be one LIZ per cell.
    \item Each cell must have enough inputs/outputs so that the ions can move to other cells.
    \item The number of ions per cell must be large enough so that significant shuttling and gates can be executed.
    \item The number of ions per cell must be small enough for the generation of a shuttling sequence to be a simple task.
\end{itemize}

The last two principles require a trade-off as they are contradictory. A small cell with a small number of ions means that the intra-cell shuttling can be easily studied and that a shuttling algorithm can be devised. This also separates the parallelization and shuttling problems as the calculation can be distributed across multiple cells. The approach proposed here is similar to classical multithreading approaches, whereby here the parallelization problem is separated from the shuttling problem, easing its resolution.

Furthermore, we believe that logical qubits implemented through multiple physical qubits may be kept together within one reservoir. This should considerably ease the generation of shuttling sequences from an algorithmic point of view. 

Four concepts of cells are presented in Fig.~\ref{fig:CellArchi}. Each cell design uses a different junction topology. The cell in Fig.~\ref{fig:CellArchi}~(a) uses Y-junctions (junction with 3 branches), that in~(b) X- and Y-junctions (4 branches), the cell in (c) is based on X-junctions with a LIZ at the center of an X-junction, and the cell in~(d) uses T-junctions (3 branches, with one branch perpendicular to the two others that are aligned).
Each cell has a reservoir where the unused ions are stored for the current gate. Each of these cells has a favored direction of ion displacements passing through the LIZ as it is left empty. The perpendicular direction uses the reservoirs. This means the existence of a direction, the LIZ direction, favorable for inter-cell displacements. This topological complication can be alleviated by a specific pattern and positioning of cells. The cell in (d) is much smaller than other cells and thus more space-efficient. The cell in~(c) appears to us to be the best, as every direction has the same cost; it is, however, also the most complex since a junction-LIZ needs to be realized.  

Patterns for tiling cells on a planar area are presented in Fig.~\ref{fig:CellArchiPatern}. As (a) and~(b) show, rotation and mirroring reduce the problem of the favored direction, but it is still present. The best solution to this problem appears to be that shown in~(c), since there is no favored direction. In (d), there is a need to efficiently link the cell in two ways; (d) leads to a larger number of cells than the other patterns, but at the cost of more difficult shuttling.

Each of the cells presented here has advantages and disadvantages. Their ease of fabrication may decide which cell might be used for future shuttled ion trap quantum computers.

\section{Conclusion}
\label{sect:Conclu}

In this paper, multiple architectures have been proposed and analyzed. It was shown that the multi-LIZ can relieve the main problem of the uni-LIZ architecture, that is the need to use a massive overhead of ion exchanges during shuttling. Another proposed architecture, the tree architecture, appears more difficult to realize, but is algorithmically far more efficient. The need for charge balancing around the LIZ and junction realization is still a complexity challenge in the design of segmented trap ion quantum computers and must be accounted for in architecture generation and shuttling algorithms.

Finally, complex architectures created from cells as building blocks have been presented. Each of these architectures needs the development of specific junctions, but such hurdles might be overcome in the shorter or middle term if ion trap quantum computers are to be scaled up. The advantage of further studying such architectures is well worth the effort, as it will increase the effectiveness of shuttling by reducing its complexity.

\section*{Acknowledgements}

J.D. acknowledges financial support from the QSciTech training program funded by the Collaborative Research and Training Experience (CREATE) program of the Natural Sciences and Engineering Research Council of Canada (NSERC). Y.B.L. acknowledges the support of the Canada First Research Excellence Fund (CFREF) through the Institut quantique at Universit\'{e} de Sherbrooke. U.P. and F.S.K. acknowledge funding by the German Bundesministerium für Forschung, Technologie und Raumfahrt (BMFTR) within the projects SYNQ and ATIQ, and by the Deutsche Forschungsgemeinschaft (DFG) within the project comfortQC in the SPP2514. The research is based upon work supported by the Office of the Director of National Intelligence (ODNI), Intelligence Advanced Research Projects Activity (IARPA), via the U.S. Army Research Office grant W911NF-16-1-0070. The views and conclusions contained herein are those of the authors and should not be interpreted as necessarily representing the official policies or endorsements, either expressed or implied, of the ODNI, IARPA, or the U.S. Government. The U.S. Government is authorized to reproduce and distribute reprints for Governmental purposes notwithstanding any copyright annotation thereon. Any opinions, findings, conclusions or recommendations expressed in this material are those of the author(s) and do not necessarily reflect the view of the U.S. Army Research Office.

%\bibliographystyle{unsrt}
%\bibliography{reference.bib}

\renewbibmacro{in:}{}
\printbibliography

%\EOD

\end{document}